\documentclass[sigconf]{acmart}

\renewcommand\footnotetextcopyrightpermission[1]{} 

\PassOptionsToPackage{usenames,dvipsnames}{xcolor}
\usepackage[T1]{fontenc}
\usepackage[utf8]{inputenc}
\usepackage[english]{babel}
\usepackage{etex}
\usepackage{amsmath}
\usepackage{mymacros}
\usepackage{thmtools}
\usepackage{comment}
\usepackage{algorithm}
\usepackage{algorithmicx}
\usepackage{algpseudocode}
\usepackage{pgfplots}
\pgfplotsset{compat=1.14}
\usepgfplotslibrary{fillbetween}
\allowdisplaybreaks
\usepackage[dvipsnames]{xcolor}

\begin{document}
\title{Option Hedging with Risk Averse Reinforcement Learning}

\author{Edoardo Vittori$^{1,2}$, Michele Trapletti$^{1}$, Marcello Restelli$^{2}$}
\email{{edoardo.vittori, michele.trapletti}@intesasanpaolo.com}
\email{marcello.restelli@polimi.it}
\affiliation{%
   \institution{$^1$Intesa Sanpaolo, $^2$Politecnico di Milano}
 }

\begin{abstract}

In this paper we show how risk-averse reinforcement learning can be used to hedge options. We apply a state-of-the-art risk-averse algorithm: Trust Region Volatility Optimization (TRVO) to a vanilla option hedging environment, considering realistic factors such as discrete time and transaction costs.
Realism makes the problem twofold: the agent must both minimize volatility and contain transaction costs, these tasks usually being in competition.
We use the algorithm to train a sheaf of agents each characterized by a different risk aversion, so to be able to span an efficient frontier on the volatility-p\&l space.
The results show that the derived hedging strategy not only outperforms the Black \& Scholes delta hedge, but is also extremely robust and flexible, as it can efficiently hedge options with different characteristics and work on markets with different behaviors than what was used in training.

\end{abstract}


\keywords{Deep Hedging, Reinforcement Learning, Transaction Costs}

\maketitle
\pagestyle{plain} 
\section{Introduction}
\label{sec:intro}

Vanilla options, contracts that offer the buyer the right to buy or sell a certain amount (the option's notional) of the underlying asset at a predefined price (the strike) at a certain future time (the maturity), are a fundamental building block in the derivatives business. They offer an investor the opportunity to take advantage of a movement in the price of an asset aligned with her market view, avoiding the risk of losing money due to the chance that the asset price moves instead in the opposite way, at a cost, the option premium, completely defined at inception.

In the options market a crucial role is played by traders
who quote both the buy and sell prices of an option. They are usually referred to as sell-side or market-making traders and generally cover a large range of options on very few underlying instruments. When a market-maker succeeds to sell {\it and} buy the same amount of the same option, she makes a profit, which is her main aim, equal to the difference between the two premia, and the financial risk is perfectly offset since the value of the portfolio realized by the two trades does not depend on the underlying asset. However, most of the time, requests are not symmetric, leaving an open risk position that must be closed (i.e., hedged).

Option pricing and hedging builds on the standard Black \& Scholes (B\&S)~\cite{black1973pricing} model which is based on a strong set of assumptions that tend to be unrealistic \cite{yalincak2012criticism}: hedging is assumed to be cost-less and continuous. Due to this, the hedging process, which  usually consists of a purchase or sale of the option’s underlying in a quantity based on the first-order derivative of the B\&S price (known as delta), must be adjusted with the trader's experience in order to both reduce the risks and contain hedging costs.

In this paper we focus on the option hedging problem in a realistic environment where we exploit the power of Reinforcement Learning (RL). In some sense we aim at replicating and hopefully improving, in an automatic way, the trader's experience of containing both risk and hedging costs.
While there is an extensive literature on both option hedging~\cite{hull2017optimal} and reinforcement learning~\cite{Suttonbook:1998}, there are very few works on the combined topics, the main ones being~\cite{kolm2019dynamic,cao19, buehler2019deep,halperin2017qlbs}, which we will analyze in Section~\ref{sec:related}.

Here we implement a robust tool capable of providing the trader with a hedging signal more accurate than the delta hedge, as it is optimized in a realistic environment, with discrete time and transaction costs. We achieve this result through the use of risk-averse RL by applying TRVO~\cite{bisi2019risk}, an algorithm capable of optimizing together the hedging (i.e. risk reduction) and p\&l objectives.

By controlling the risk-aversion parameter, we are able to create a Pareto frontier in the volatility-p\&l space, which strictly dominates the delta hedge since it performs better both in terms of variance and in terms of p\&l.
Having trained a sheaf of agents, each characterized by a different risk aversion, we reduce the job  of the trader to simply deciding on which point of the frontier to place herself.
We also show that the trained agents are robust since they can efficiently hedge options with different characteristics and work on markets which behave differently than what was used in training.
These experimental results are the main contribution of this work.

The paper is structured as follows: in Section~\ref{sec:delta} we present the hedging of a vanilla option using the B\&S model and subsequently explain how the hedging environment can be embedded to work in a reinforcement learning context. For simplicity, we restrict our study to a long position in call options of unitary notional (extensions to put options or to short positions or to a generic notional are trivial).
In Section~\ref{sec:RL} we describe the chosen reinforcement learning algorithm. In Section~\ref{sec:experiments} we present and evaluate the empirical performance. In Section~\ref{sec:related} we compare with the current literature and in Section~\ref{sec:conclusions} we present our conclusions and outlook.

\section{Delta hedging}
\label{sec:delta}
In this section we describe the Black \& Scholes pricing framework. It is the main option model used in practice, and thus the benchmark considered in this paper. There are two things to model: the underlying and the deriving option price. 
In the B\&S framework, the underlying behaves as Geometric Brownian Motion (GBM), thus let $S_t$ be the underlying at time $t$, then it can be described as:
\begin{equation*}
    \mathrm{d}S_t = \mu S_t \mathrm{d}t + \sigma S_t \mathrm{d}W_t,
\end{equation*}
where $W_t$ is Brownian motion, $\mu$ the drift (which we assume to be 0 throughout the paper without loss of generality) and $\sigma$ the volatility.
For an initial value $S_0$, the SDE has the analytic solution:
\begin{equation*}
    S_t=S_0\exp\left(\left(\mu-\frac{\sigma^2}{2}\right)t+\sigma W_t\right),
\end{equation*}
where: $W_{t+u}-W_t \sim \mathcal{N}(0,u) = \mathcal{N}(0,1)\times \sqrt{u}$, $\mathcal{N}$ being the normal distribution.
Let $C_t$ be call option price at time $t$, $T$ the time \textit{of} maturity, $T-t$ the Time \textit{To} Maturity (TTM), $K$ the strike price, $\sigma$ and $\mu$ coincide with those of the GBM. The B\&S call price $C_t$ is: 
\begin{align*}
    &C_t(S_t) =  \Phi(d_t)S_t-\Phi(e_t)Ke^{\mu(T-t)},\\ 
    &d_t = \frac{1}{\sigma \sqrt{T-t}}\left[\ln\left(\frac{S_t}{K}\right)+\left(\mu+\frac{\sigma^2}{2}\right)(T-t)\right],\\
    &e_t = d_t - \sigma\sqrt{T-t},
\end{align*}
where $\Phi$ is the cumulative distribution function of the standard normal distribution. We introduce $\frac{\partial C_t}{\partial S}$, which is known as the option delta and for our position (a long call of unitary notional) is bounded between 0 and 1. In particular when $T-t$ is relatively small and $\frac{S_t}{K}\ll 1$, $\frac{\partial C_t}{\partial S} \rightarrow 0$ and $C_t \rightarrow 0$; instead if $\frac{S_t}{K}\gg 1$, $\frac{\partial C_t}{\partial S} \rightarrow 1$ and $C_t\rightarrow S_t$. 
A trader who has a long position in a call option will endure a profit swing of $C_{t+k}(S_{t+k})- C_t(S_{t})$ for a time-lag of $k$.
A delta hedge is a strategy to limit this profit movement by buying or selling a certain quantity of the underlying, call this function $h(\frac{\partial C_t}{\partial S_t}, \mathcal{E})$, which depends on the delta and the trader's experience $\mathcal{E}$; we will refer to it as just $h_t$ for ease of notation.

The profit in one timestep of a trader who bought a call option and making a delta hedge can be calculated as $\rho_{t+k}= C_{t+k}(S_{t+k})- C_t(S_{t}) - h_t\times(S_{t+k} - S_{t}).$
Now assume that we replicate the delta exactly, so $h_t = \frac{\partial C_t}{\partial S_t}$ and there is no experience into play. Then the B\&S model assures a zero profit in the continuous limit ($k\rightarrow 0$) and in the absence of transaction costs.

From now on, we consider $k>0$, in particular we take as a reference point 5 evenly spaced rebalances of $h_t$ per day. Another point of realism is adding transaction costs for the underlying hedging instrument, which we define as in~\cite{kolm2019dynamic}:
\begin{equation}
\label{eq:costs}
    c(n) = \text{Ticksize} \times (|n|+0.01n^2),
\end{equation}
where $n = h_{t}-h_{t-1}$. So the $\rho$ becomes:
\begin{equation}
  \rho_{t+k}= C_{t+k}(S_{t+k})- C_t(S_{t}) - h_t\times(S_{t+k} - S_{t}) - c(n).
    \label{eq:pl}
\end{equation}
With $k>0$ and costs, $ h_t = \frac{\partial C_t}{\partial S_t}$ ceases to be the optimal solution.

\section{Learning to hedge}
\label{sec:RL}
In this section we give a brief introduction to reinforcement learning, focusing on policy search algorithms. After defining the basics, we explain how to embed the hedging environment in this framework, we conclude by discussing risk-aversion and safety.

\subsection{Reinforcement Learning}
A discrete-time Markov Decision Process (MDP) is defined as a tuple $\langle\Sspace,\Aspace, \mathcal{P}, \mathcal{R}, \gamma, \mu\rangle$, where $\Sspace$ is the state space, $\Aspace$ the (continuous) action space, $\mathcal{P}(\cdot|s,a)$ is a Markovian transition model that assigns to each state-action pair $(s,a)$ the probability of reaching the next state $s'$, $\mathcal{R}(s,a)$ is a bounded reward function, $\gamma\in[0,1)$ is the discount factor, and $\mu$ is the distribution of the initial state. The policy of an agent is characterized by $\pi(\cdot|s)$, which defines for each state $s$ an action with a probability distribution over the action space.
We consider infinite-horizon problems in which future rewards are exponentially discounted with~$\gamma$.  Following a trajectory $\tau \coloneqq (s_0, a_0, s_1, a_1, s_2, a_2, ...$), let the returns be defined as the discounted cumulative reward:
\begin{equation}
G = \sum_{t=0}^\infty \gamma^t \mathcal{R}(s_t,a_t).
\label{eq:returns}
\end{equation}
For each state $s$ and action $a$ the action-value function is defined as:
\begin{equation*}
    Q_\pi(s,a) \coloneqq \EV_{\substack{s_{t+1}\sim \mathcal{P}(\cdot|s_{t},a_{t})\\a_{t}\sim\pi(\cdot|s_{t})}}\left[\sum_{t=0}^\infty \gamma^t \mathcal{R}(s_t,a_t)|s_0 = s, a_0 = a\right],
    \label{eq:Q_fun}
\end{equation*}
which can be recursively defined by the Bellman equation:
\begin{equation}
        Q_\pi(s,a) = \mathcal{R}(s,a) + \gamma \EV_{\substack{s'\sim \mathcal{P}(\cdot|s,a)\\a'\sim\pi(\cdot|s')}}\big[Q_\pi(s',a')\big].
        \label{eq:bellman}
\end{equation}
The typical RL objective $J_\pi$ is defined as
\begin{align}
    J_\pi & \coloneqq
    \hspace{-19pt}
    \EV_{\substack{s_0\sim \mu \\a_t\sim\pi(\cdot|s_t)\\s_{t+1}\sim \mathcal{P}(\cdot|s_{t},a_{t})}}
    \hspace{-9pt}
    \left[\sum_{t=0}^\infty \gamma^t \mathcal{R}(s_t,a_t)\right] 
    = \EV_{\substack{s_0 \sim \mu\\ a\sim\pi(\cdot|s)}}\left[ Q(s,a)\right].
    \label{eq:objective}
\end{align}
This objective can be maximized in two main ways. The first is by learning  the optimal action-value function $Q^*$ for each state and action, thus the optimal policy is: $\pi^*(a|s) = \mathrm{argmax}_{a} Q^*(s,a)$. These algorithms are called value-based \cite{Suttonbook:1998}, and are used in~\cite{kolm2019dynamic,cao19}. This type of algorithm can become impractical in a hedging environment where both states and actions are continuous. In fact, it is necessary to approximate the value function, and it has been shown that even in relatively simple cases, such algorithms fail to converge \cite{baird1999gradient}. 

The other family is instead policy search methods \cite{deisenroth2013survey}, which optimize the objective by searching directly in the policy space. They can easily handle continuous actions, learn stochastic policies in partially observable, non-Markovian environments and are robust when working in datasets with large amounts of noise \cite{moody2001learning}. 

Furthermore, it is possible to build algorithms with safety guarantees~\cite{kakade2002approximately, pirotta2013adaptive, schulman2015trust, schulman2017proximal, papini2017adaptive} (i.e., the policy ensures an improvement at each update). We focused on the algorithm which has given some of the best experimental results in the field \cite{OpenAI_dota, Heess2017Emergence}: Trust Region Policy Optimization (TRPO) \cite{schulman2015trust}. It is a practical algorithm, deriving from an approximation of a theoretically justified procedure which optimizes a local approximation to the expected return of the policy with a KL divergence constraint.

\subsection{Embedding in a Markov Decision Process}
\label{ssec:embedding}
In this section, we explain how the option hedging framework can be embedded in an MDP:
\begin{itemize}
    \item the action $a_t$ is the current hedge portfolio, replaces $h_t$,
    \item the state $s_t = (S_t,C_t,\frac{\partial C_t}{\partial S_t},a_{t-1})$,
    \item the reward $\mathcal{R}(s_t,a_t) = f(\rho_t)$, with $\rho_t$ as in Equation~(\ref{eq:pl}).
\end{itemize}
The above formulation is similar to the one used in \cite{kolm2019dynamic}, and is called \textit{accounting formulation} in \cite{cao19}. There is another possible MDP formulation, defined in \cite{cao19} as \textit{cash flow formulation} and used in \cite{buehler2019deep}:
\begin{itemize}
    \item the action $a_t$ is the current hedge portfolio,
    \item the state $s_t = (S_t,T-t,a_{t-1})$,
    \item the reward $\mathcal{R}(s_t,a_t) = f(\rho^{CF}_t)$ where 
    \begin{equation*}
    \rho^{CF}_t= a_t(S_{t+1} - S_{t}) -c(n)+CF_t,
    \end{equation*}
\end{itemize}
where $CF_t$ represents the cash flows of the derivative which is being hedged. In the case of options, this means the initial premium spent to buy the option and the payoff at expiry or $(S_T-K)^+$. The major difference between the two approaches is that, in the cash flow case, it is not necessary to have a pricing model for the option. In \cite{kolm2019dynamic}, the option price (and delta) is not included in the state, but remains necessary to calculate the reward which is defined following the accounting formulation.

In this paper we consider the accounting formulation for several reasons. First of all, the fact that the state includes the option price and delta makes the policy learned robust to changes in the option characteristics, as we will see in Section~\ref{sec:strengths}. A second reason is that the agent is capable of learning more rapidly as it does not need to implicitly reconstruct the B\&S pricing. Finally, in any type of real trading environment, calculating the option price and delta is a standard and already implemented. 

We would like to emphasize that even though we assume that the actions do not impact market prices, it is necessary to use RL as the current portfolio is included in the state. Without this formulation, the agent would not be able to plan ahead and thus minimize costs.
\subsection{Risk Averse and Safe Reinforcement Learning}
\label{ssec:riskaverse}
 As we can see from Equation~(\ref{eq:objective}), if $\mathcal{R}_t = \rho_t$, then the objective means maximizing the expected cumulative $\rho_t$. Maximizing this quantity is not the correct objective for this type of problem, in fact in an ideal B\&S model, this quantity is as close as possible to zero, which translates in optimizing a risk-averse objective. There are several possible approaches at this point: we could keep the same algorithm simply by considering $\mathcal{R}_t=-|\rho_t|$ or $\mathcal{R}_t = \min(\rho_t,0)$,
 however, we have a further objective, since we want to also maximize the $\rho_t$; optimally, we would like to decide how much to focus on maximizing the cumulated rewards and how much on minimizing the risk. We can do this by modifying the reward as:
 $\mathcal{R}_t = \rho_t-\lambda {\rho_t}^2$ as in~\cite{kolm2019dynamic}. In this case $\lambda$ is a term that balances the risk-reward trade-off, but the term $\rho_t^2$ is an approximation of a variance term and thus could converge to suboptimal points. Finally, the reward can be modified as $\mathcal{R}_t = e^{-\lambda \rho_t}$, thus transforming the reward in a utility function~\cite{shen_risk-averse_2014}, with peculiar properties from a mathematical perspective (see also~\cite{buehler2019deep}). Unfortunately, we experimented two negative behaviors: for high $\lambda$ values the rewards explode causing the RL algorithm to crash, thus making the generation of highly risk averse policies a very difficult task from a numerical perspective; for really low values of lambda (trying to obtain a risk neutral behavior) the reward essentially becomes constant and there is no learning. 
 
 Another possibility is to change the objective, but this means modifying the actual learning algorithm. In particular, using the definition of returns (Equation~(\ref{eq:returns})), it is possible to define the objectives in the equations below. In order to simplify notation, and unless specified otherwise, the expected values and variances are calculated over the initial state $s_0\sim \mu$, and the trajectory given by sequence of states $s_{t+1}\sim \mathcal{P}(\cdot|s_{t},a_{t})$ and actions $a_t\sim\pi(\cdot|s_t)$:
 \begin{align}
    MV_\pi & \coloneqq \EV[G] - \lambda \Var[G]
    \label{eq:mean-variance}\\
    SR_\pi & \coloneqq \EV[G]/\sqrt{\Var[G]}.
    \label{eq:ra_objective}
\end{align}
 $MV$ or \textit{mean-variance} was introduced in \cite{sobel_variance_1982, Tamar_variance, tamar_variance_2013, prashanth2016variance}, while the $SR$ or \textit{Sharpe ratio} was defined in \cite{moody2001learning}. We will refer to the variance term as \textit{return-variance}, it represents the variance of the cumulated rewards.
 Some techniques consider CVaR, or more generalized risk measures called coherent risk measures and defined in \cite{tamar_policy_2015,Morimura2010NonparametricRD, chow2017risk}.

 The problem with most of the previous approaches is that the learning algorithm is not effective for the complex environment analyzed in this paper, for this reason we considered the RL algorithms that have given the best experimental results. 
 The final choice was TRPO as it has a risk averse variant: TRVO defined in~\cite{bisi2019risk}.
 The objective is $\eta_{\pi} = J_{\pi}+\lambda \nu_{\pi}^2$, where:
\begin{equation}
    \nu^2_\pi \coloneqq \EV\left[\sum_{t=0}^\infty \gamma^t \left(\mathcal{R}(s_t,a_t)-J_\pi\right)^2\right].
    \label{eq:reward_vola} 
\end{equation}
One of the interesting properties of this risk metric, which we will refer to as \textit{reward-variance}\footnote{To be consistent with a financial environment, We refer to this metric as reward variance, even though in the original paper it is called reward volatility.}, is that it bounds the return-variance term in Equation~(\ref{eq:mean-variance}) ($\Var[G]$). We would like to draw the reader's attention to the meaning of this reward-volatility term: it is minimizing the variations between one step and the next, in contrast to the return-variance which is minimizing the variance at the end of each path. 
TRVO naturally converges to TRPO if $\lambda \rightarrow 0$.

As a final remark we stress the fact that TRVO has good convergence features for essentially any value of $\lambda$, offering the chance of training agents with very different risk aversions.

\section{Experiments}
\label{sec:experiments}

In this section we present some of the experimental results. 
After introducing the financial environment in Section~\ref{ssec:fin_env}, we consider in Section~\ref{ssec:no_costs} option hedging in discretized time without hedging costs; then in Section~\ref{sec:experiment_std_cost} we introduce transaction costs. Finally, in Section~\ref{sec:strengths} we stress our models by testing them on options with different strikes, different volatility, and even portfolios of options.
Training is performed on 10,000 scenarios, with $\sim$2,000 episodes where each batch has $\sim$120,000 steps, the discount factor is $0.999$, max KL is $0.001$. The policy is a neural network with 2 hidden layers with 64 neurons. Testing is an average of 2,000 out-of-sample scenarios. All figures present the results of those 2,000 scenarios unless specified otherwise.

\subsection{Financial environment}
\label{ssec:fin_env}
We consider a single underlying, generated with GBM as explained in Section~\ref{sec:delta}. There are 5 prices per day. The option has unitary notional and is At The Money (ATM) with a 60 day maturity \footnote{We have chosen an ATM option because it is the most traded type and generates the most interesting delta behavior. Nevertheless, in Section~\ref{sec:strengths} we show that an agent trained on ATM options is able to hedge options with different moneyness.} . The starting price of the underlying is 100 and the annual volatility is 20\%: approximately that of an equity option in normal market conditions. With these characteristics, the option price value at inception is $\sim 3.24$ (also referred to as premium) and the corresponding delta $\sim 0.5$. 305 timesteps means 61 days, where day 61 is the day the option expires. On the last day, small movements of the underlying around the strike price can cause the delta to jump between 0 and 1, this is called pin risk. Finally, we simplify a zero financing cost.
In this section, we will refer to $\nu$ (Equation~(\ref{eq:reward_vola})) as reward volatility, p\&l := $\sum_t \rho_t$ and p\&l volatility is $\sigma :=\sqrt{\Var [\mathrm{p\&l}]}$.

\subsection{Training without transaction costs}
\label{ssec:no_costs}
\begin{figure}[t]
    \centering
    \includegraphics[width=.45\textwidth]{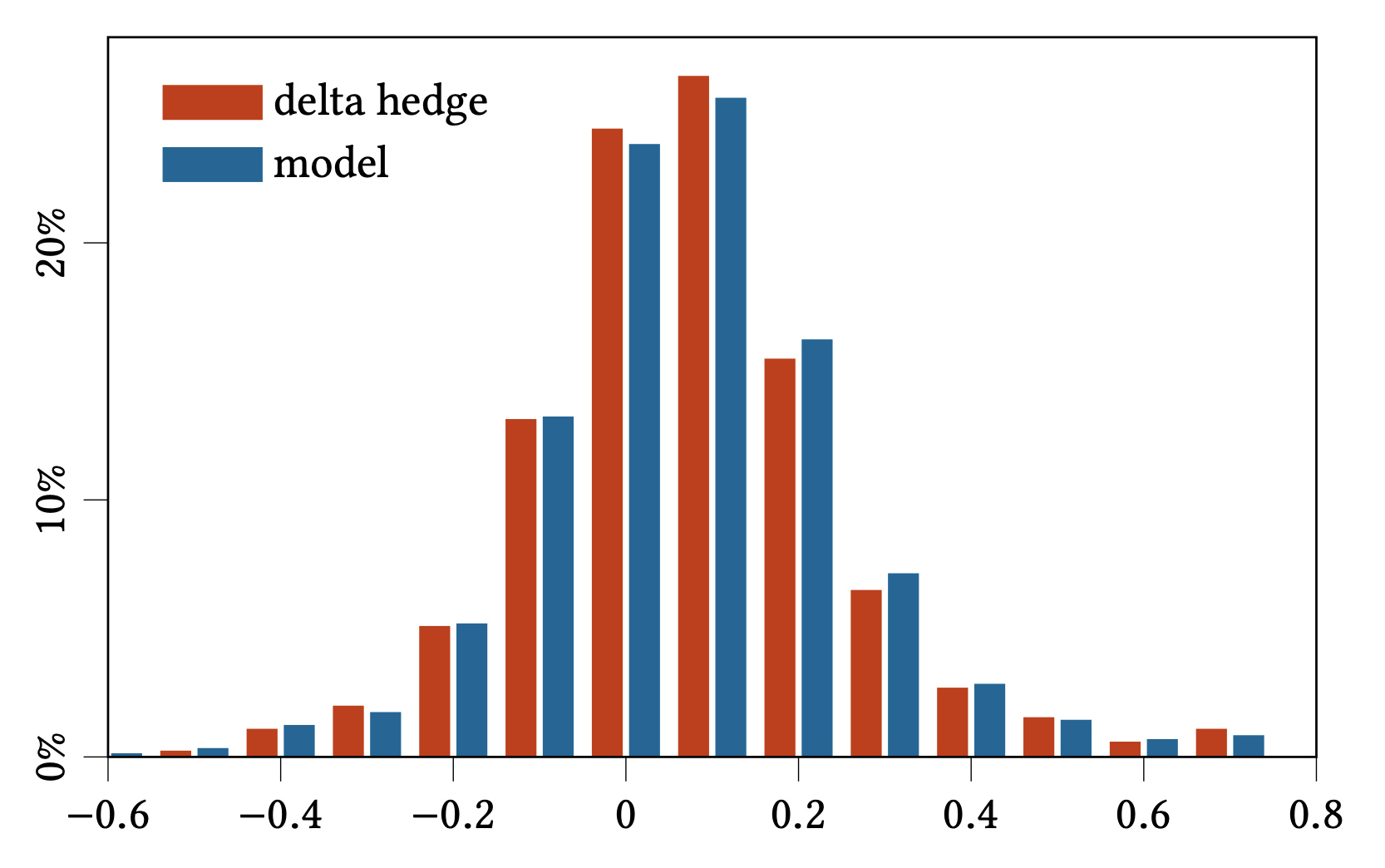}
    \caption{P\&l distribution without transaction costs.}
    \label{fig:no_cost_PNL_dist}
\end{figure}

\begin{figure}[t]
    \centering
    \includegraphics[width=.45\textwidth]{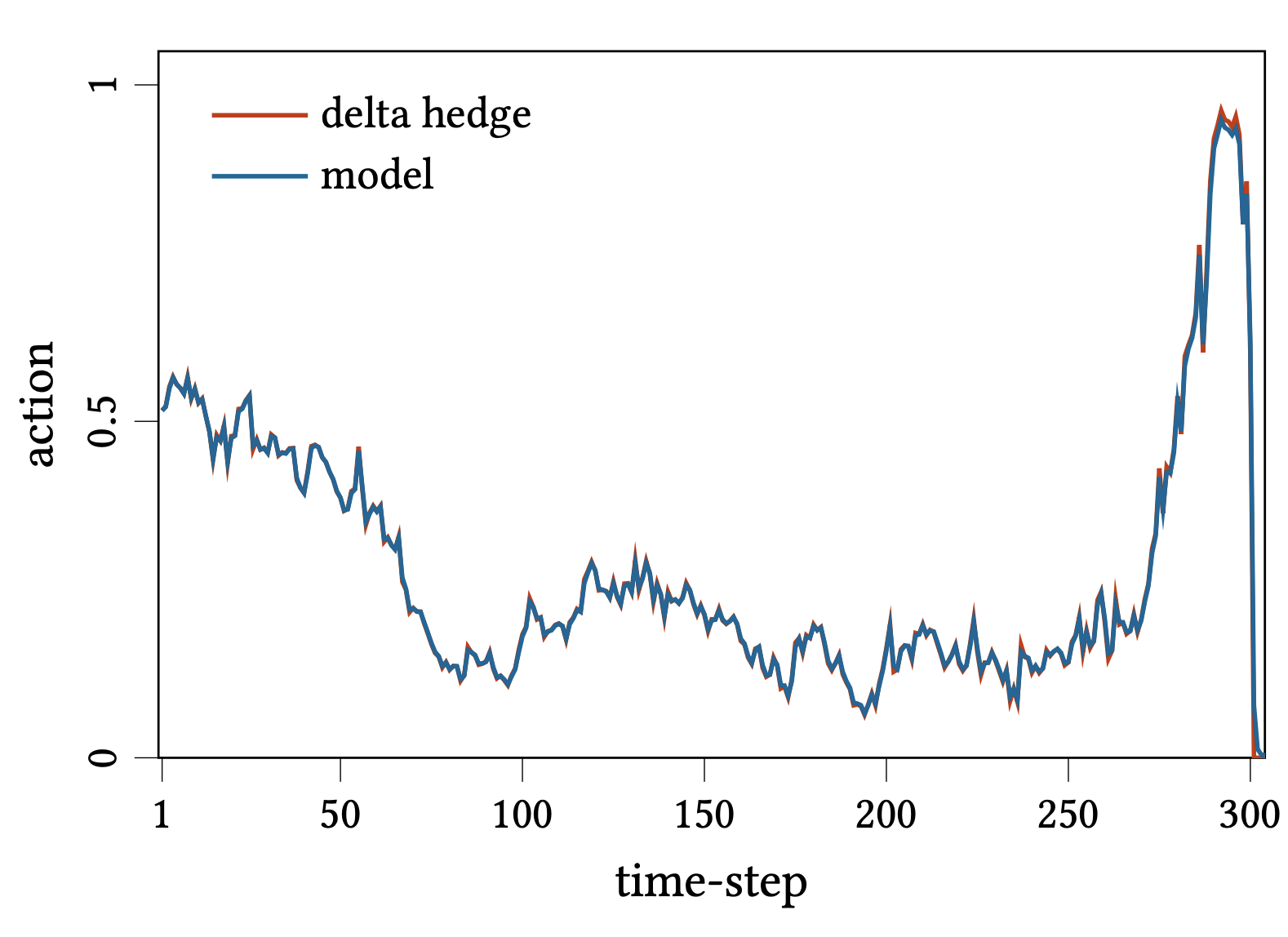}
    \caption{Single scenario, no transaction costs with also an example of pin risk handled appropriately by the agent.} 
    \label{fig:no_cost_delta_vs_agent}
\end{figure}

In this section we analyze the performance of TRVO on a cost free environment and compare it with the delta hedge.
As we can see in Figure~\ref{fig:no_cost_PNL_dist}, time discretization injects some volatility in the delta hedge (red bars), but the generated wealth is very small with respect to the option premium ($\sim 3.24$).
In fact, the delta hedge remains optimal and the agent learns to reproduce it (as we can see in Figure~\ref{fig:no_cost_delta_vs_agent}). It is clear that the hedging strategy is able to exploit the change in delta typical of a long option position, in order to compensate for the time decay of the option premium.
The presence of a risk aversion factor, even if slightly weighted, forces the algorithm to reduce the volatility as much as possible, thus there is no frontier and everything condenses to the delta hedge.
This feature is of course specific to the functional form of the utility function we have chosen, in some sense we recover the results of \cite{buehler2019deep} in the cost free environment assuming a 50\%-CVaR strategy (see also \cite{halperin2017qlbs,halperin2019qlbs} for similar works).
\subsection{Training with transaction costs}
\label{sec:experiment_std_cost}
\begin{figure}[t]
    \centering
    \includegraphics[width=.45\textwidth]{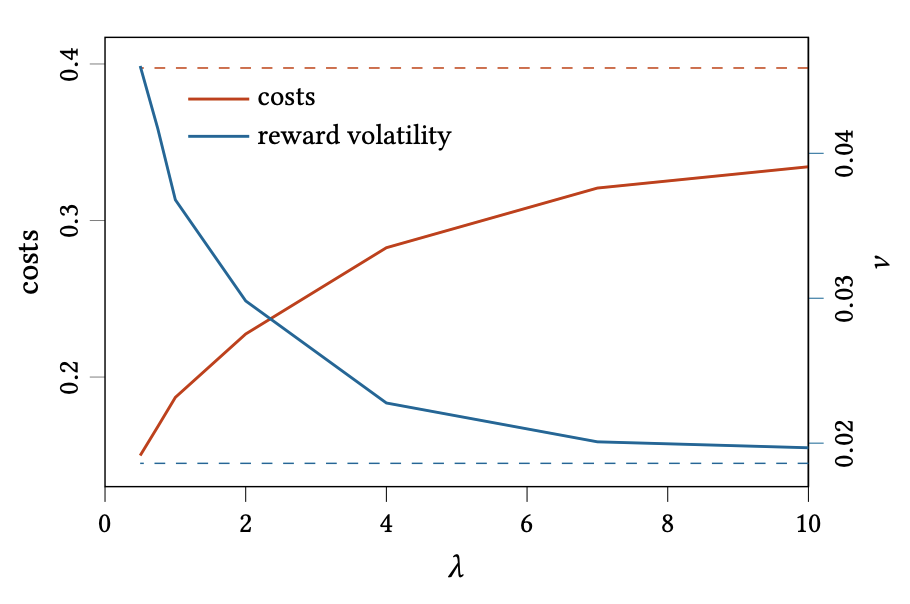}
    \caption{Hedging costs (red) and reward volatility (blue) experienced by the TRVO agent, as functions of the training risk aversion $\lambda$. The dotted lines represent the hedging cost (red) and reward volatility (blue) of the delta hedge.  Each point is calculated on a single scenario, the same of Figure~\ref{fig:no_cost_delta_vs_agent}.} 
    \label{fig:sigma_wealth}
\end{figure}
In this section, we consider costs along the line of~\cite{kolm2019dynamic} as specified in Equation~(\ref{eq:costs}) with $\text{tick size} = 0.05$.
The choice of the tick size is such that the costs replicate those of listed stocks (the
Euro Stoxx 50 future and FTSE MIB futures, renormalizing the underlying to 100, have a rescaled tick size $\sim 0.05$. A more liquid index such as the S\&P 500 ``mini'' futures contract, has a typical tick size $\sim 0.01 $).
We picked risk averseness parameter of the objective $\eta = J+\lambda \nu^2$ by measuring the typical values assumed by the reward volatility and the average p\&l. For the specific environment at hand we found $0.2\lesssim\lambda\lesssim 20$ as the most interesting range.

The average of the costs generated by the delta hedge on the test set is $\sim 0.286$, which is $\sim 9\%$ of the option premium. In Figure~\ref{fig:sigma_wealth}, the costs are actually 0.4 (red dashed line), but this is because it is a specific scenario. $0.286$ is slightly higher than what is usually experienced on the market for the bid/ask of (liquid) options on liquid underlyings (which is usually $\leq 0.1$). In our opinion, the reason for this is that, as mentioned in the introduction, market-makers do not rely solely on delta hedging in the management of their books, but rather try to match demand and offer first; furthermore, pure delta hedging is a fundamental benchmark since it is easy to reproduce, but we expect market-makers to use more sophisticated management tools.

\begin{figure}[t]
    \centering
    \includegraphics[width=.45\textwidth]{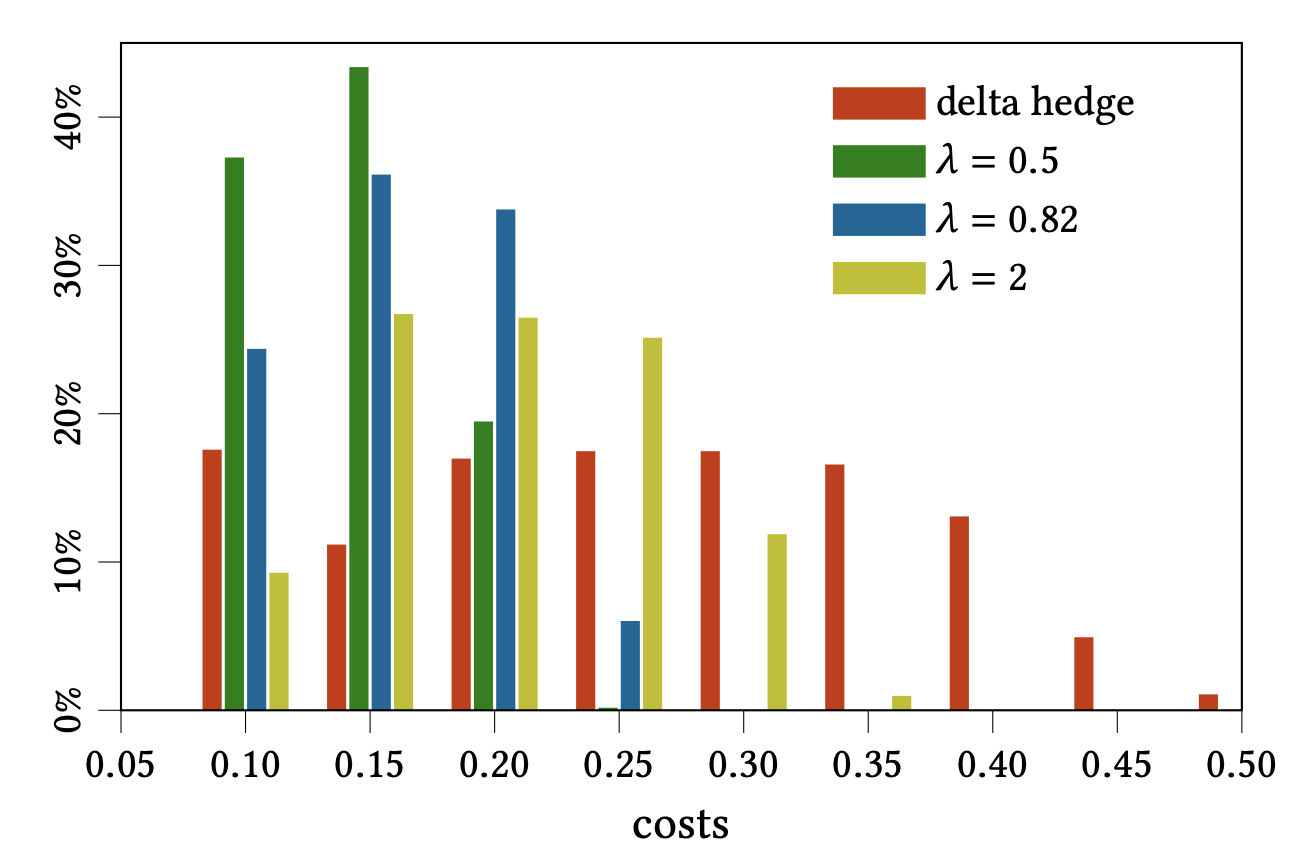}
    \caption{Hedging costs generated by delta hedge (red) and RL agent trained with different risk aversion parameters. 
      }
    \label{fig:cost_hist}
\end{figure}

In Figure~\ref{fig:sigma_wealth}, we can see that lower risk aversions generate lower costs.
This can be clearly seen also in Figure~\ref{fig:cost_hist}, where the yellow bars show that the distribution of costs generated by an agent trained with $\lambda = 2$ are much lower than those generated by the delta hedge average. This is even more evident with $\lambda = 0.5$ (green bars) and $\lambda = 0.82$ (blue bars), where costs are even lower.
Increasing the risk-aversion parameter leads the agent to behaviors more adherent to the delta hedge strategy, which is essentially recovered for $\lambda \sim 20$. This is evident from Figure~\ref{fig:sigma_wealth}, in which the costs due to the actions of the RL agent (red line) approach the costs realized by the delta hedge (dashed red line) as $\lambda$ increases.
\subsubsection{How the agent reduces hedging costs}
In light of these results, we tried to understand which strategy was chosen by the agent to reduce costs. The essence is that, even with a very low risk-aversion, the agent tries to control the p\&l volatility by mimicking the delta hedge strategy, but by delaying and reducing the action, in line with what is also described in \cite{cao19}.
This means that the agent does not act immediately when the delta spikes up (down), but waits to see if, due to market movements, the delta returns to the previous lower (higher) values. In case the delta spikes up more, surpassing the agent's ``comfort zone'', the agent covers the position. Having studied an agent trained with different risk-aversion levels, we are able to show how this comfort zone depends on the risk aversion parameter: it is very wide for lower values and very tight (or essentially zero) for higher values. 

\begin{figure}[h]
    \centering
    \includegraphics[width=.45\textwidth]{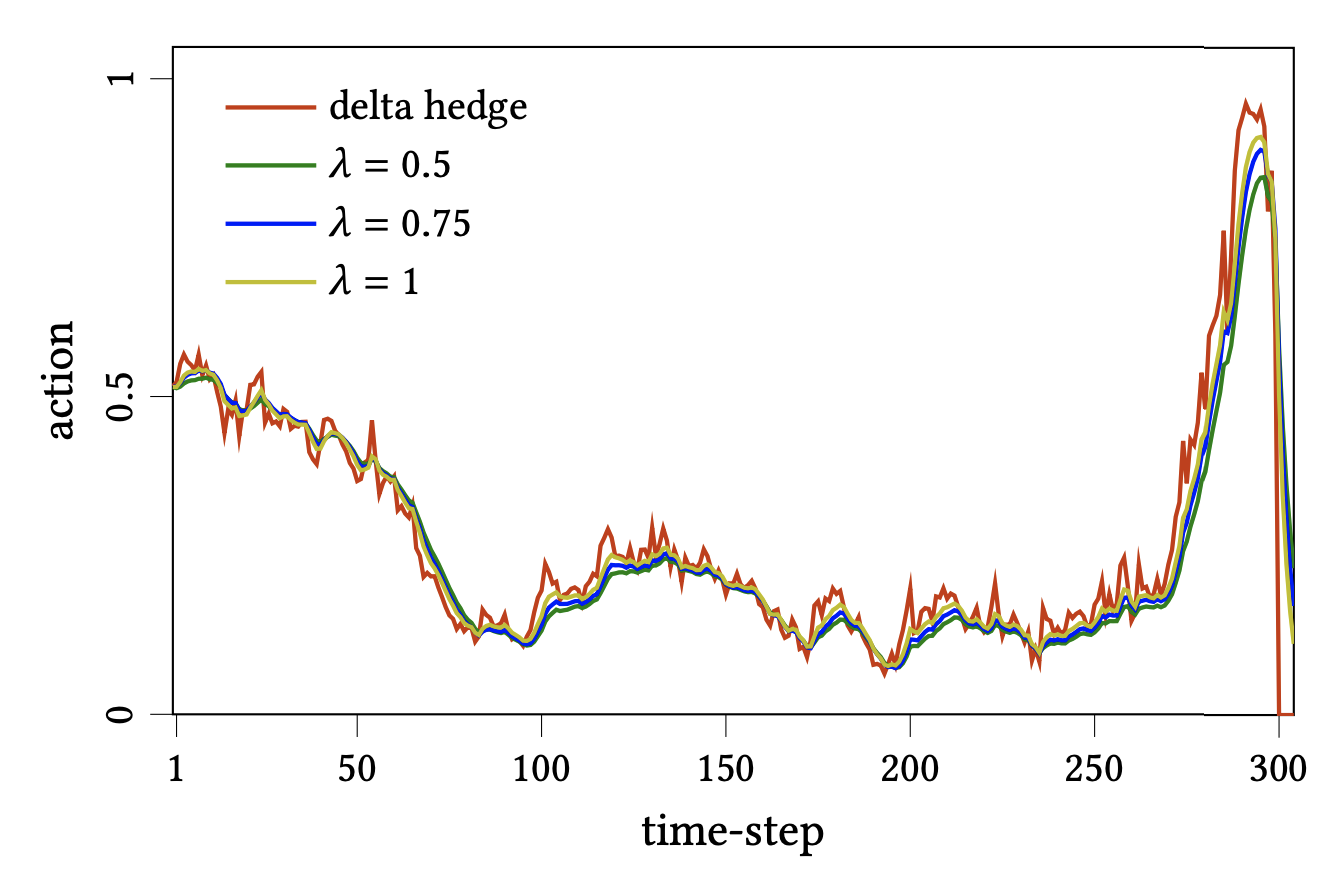}
    \caption{Comparison between the delta hedge and the agent's actions with different risk aversion parameters $\lambda$. Same scenario as figure~\ref{fig:no_cost_delta_vs_agent} but with hedging costs.} 
    \label{fig:hedge_time}
\end{figure}

This behavior is well represented in Figure~\ref{fig:hedge_time}, where the red line represents the delta hedge, while the other lines represent the action of the agent trained with different risk aversions. For lower risk-aversions, the action is smoother and expresses a significant delay w.r.t. the delta hedge. For higher values of $\lambda$, not shown for the interpretability of the figure, the agent's action is consistently  more adherent to the delta hedge, confirming the behavior which could be supposed from Figure~\ref{fig:no_cost_delta_vs_agent}.
Comparing Figures~\ref{fig:no_cost_delta_vs_agent} and~\ref{fig:hedge_time}, it is clear how the same risk aversion with different costs gives different hedging strategies.
\begin{figure}[h]
    \centering
    \includegraphics[width=.47\textwidth]{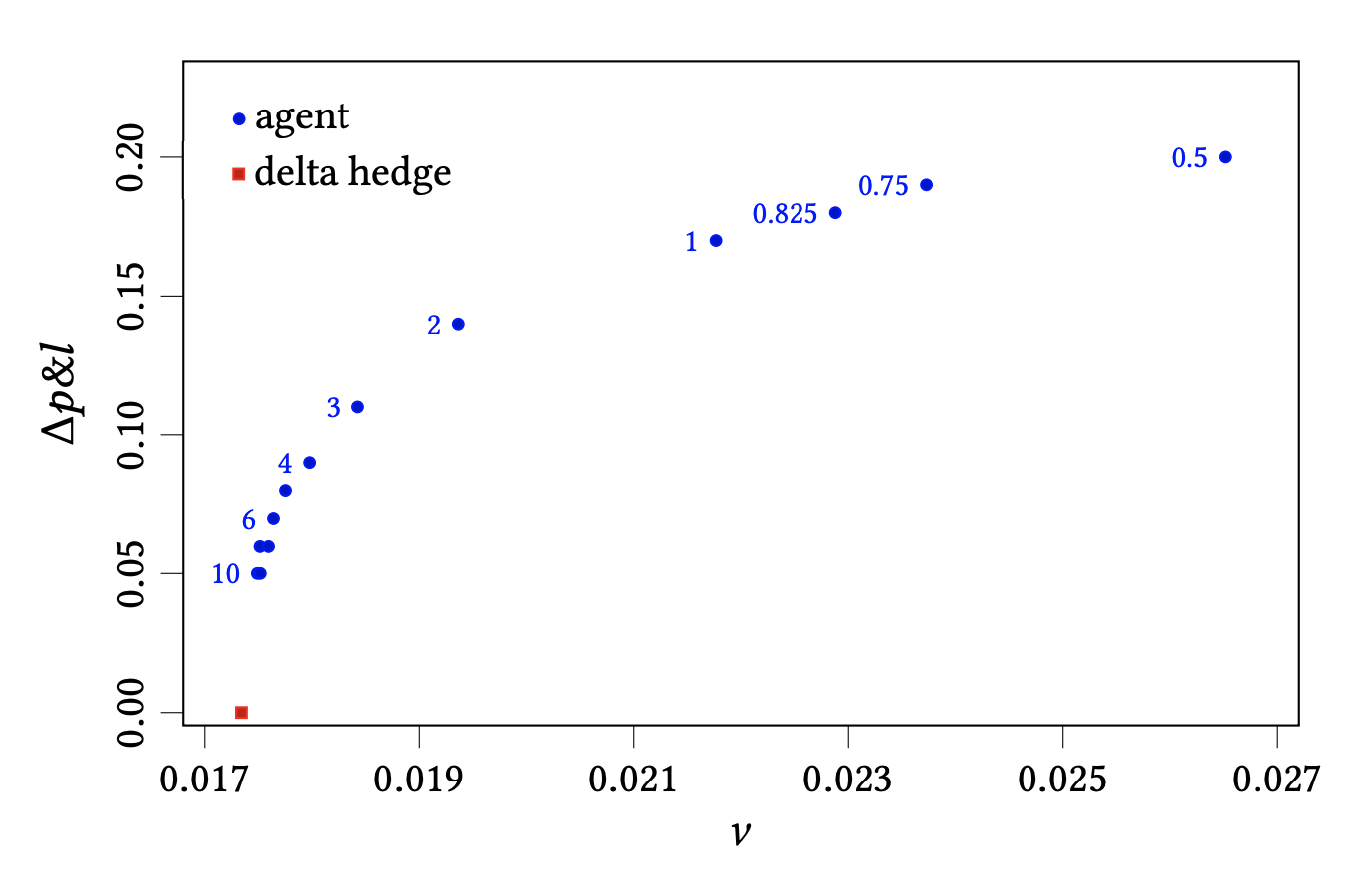}
    \caption{Efficient frontier of the TRVO agent at different risk aversions on the p\&l / reward volatility space.} 
    \label{fig:frontier0}
\end{figure}
\subsubsection{Efficient frontier}
\label{sec:eff_front}
The interplay between reward volatility reduction and cost minimization can be analyzed by observing the efficient frontier in Figure~\ref{fig:frontier0}, where we plotted a point for each value of the risk aversion parameter $\lambda $, in the $\nu$-p\&l space. The y-axis is not the pure p\&l, but the difference of the p\&l of the agent w.r.t that of the delta hedge. As expected, increasing the risk-aversion coefficient lowers the reward volatility and the p\&l.
In red, the point that shows the average wealth and reward volatility experienced by following the delta hedge strategy.
Reward volatility is a relevant risk metric to a trading strategy given that, in real life, a portfolio will experience a single scenario: if the loss is too large, the trader may be tempted, or forced, to adopt stop-loss strategies, hampering any chance of profiting from an otherwise properly trained agent. This is the main financial reason why we chose the TRVO agent.
\begin{figure}[h]
    \centering
    \includegraphics[width=.47\textwidth]{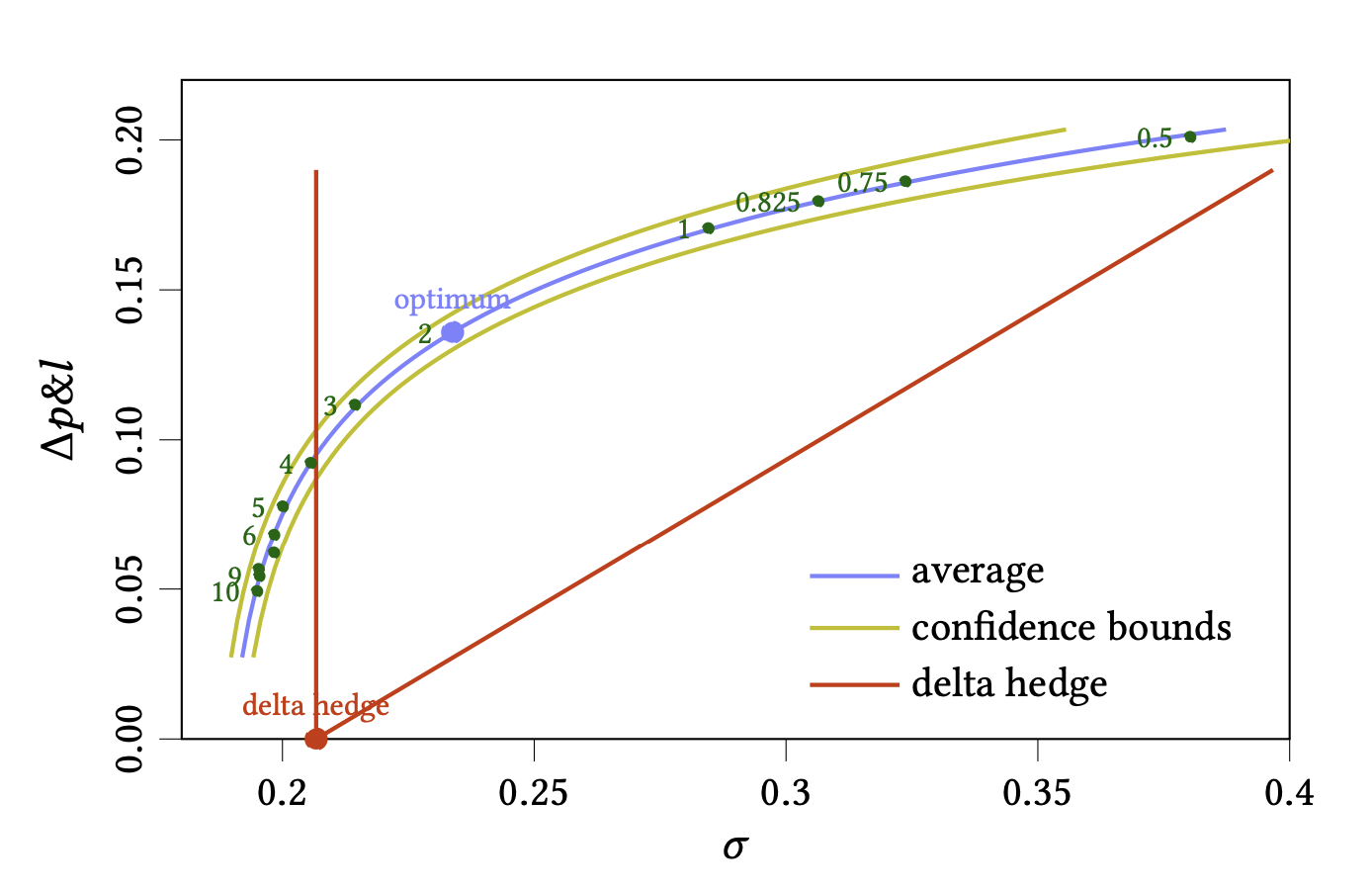}
    \caption{Efficient frontier of the TRVO agent at different risk aversions on the p\&l / p\&l volatility space.} 
    \label{fig:frontier}
\end{figure}
On the other hand, a posteriori, it is important to evaluate the performance of the hedge over the entire life of the option. This can be seen in Figure~\ref{fig:frontier} where a point in the $\sigma$-p\&l space is plotted for each value of $\lambda $. As before, the y-axis indicates the p\&l of the agent w.r.t that of the delta hedge.
In order to measure the uncertainty of the learning model, we performed 20 trainings for each value of $\lambda$.
The blue line is a logarithmic best fit of the obtained frontier, while the yellow lines provide an indication of the 1-$\sigma$ (68\%) confidence interval.
The red dot represents the delta hedge, with performance defined as being zero.
The frontier points laying on the left of the vertical red line strictly dominate the delta hedge, since the corresponding agents perform better {\it both } in terms of p\&l and in terms of volatility.
Those on the right instead, while performing even better in terms of p\&l, induce a volatility increase, thus, their relevance depends on the trader's risk aversion. As an example, a trader valuing the volatility increase on the same footing as the p\&l gain (a completely arbitrary choice, of course) will consider the whole frontier as an improvement w.r.t. the delta hedge, since it lays above the diagonal red line, indicating the region of the space where the p\&l gain is equal to the volatility increase.
That trader will consider the blue point, which has tangent line parallel to the diagonal red line and is very close to the $\lambda=2$ point, as an optimum.
We stress that, whatever risk aversion is chosen, there is a corresponding frontier point performing better than delta hedge.

\subsubsection{P\&L distribution}
\begin{figure}[h]
    \centering
    \includegraphics[width=.45\textwidth]{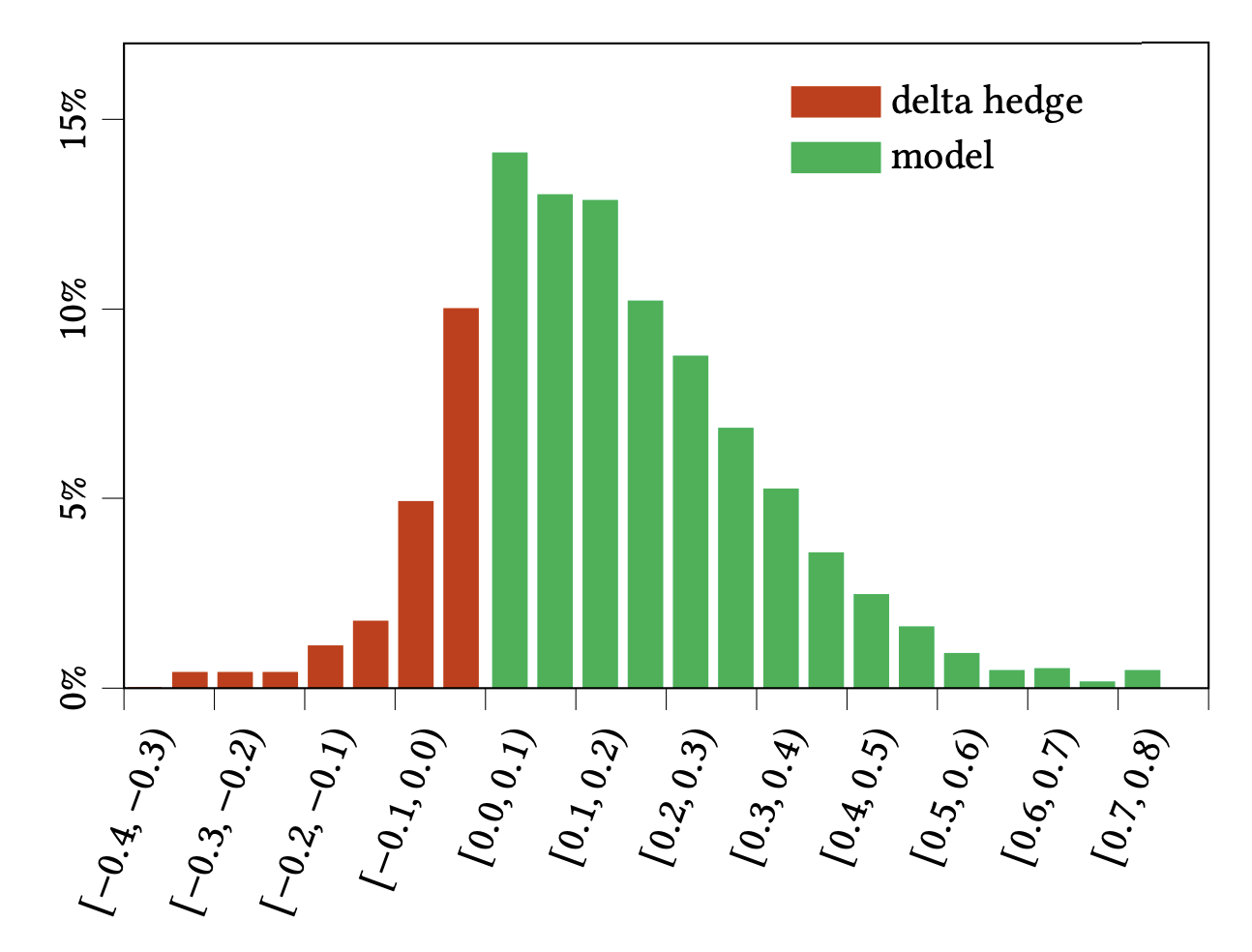}
    \caption{Distribution of the p\&l performance of a TRVO agent with $\lambda=2$ over the delta hedge.}
    \label{fig:wealth_hist}
\end{figure}

One may wonder about the statistical significance of a gain of $\sim$ 0.15 with respect to the delta hedge if the p\&l-volatility is of the same order of magnitude. We believe this is the case, and support our claim by showing, in Figure~\ref{fig:wealth_hist}, how the agent performance is distributed in the case $\lambda=2$, where the green (red) bars show the distribution of scenarios in which the agent performs better (worse) than the delta hedge. Assuming that a performance better than that of the delta hedge {\it is not always ensured}, one has to observe that in 81\% of the scenarios the performance is superior and that the average superior performance, which is 0.15, is more than ten times the average of the worse performance, which is -0.014, and even more of the absolute value of the 5-th percentile of the distribution, which is -0.09. The t-test on the data used for the histogram gives more than a 99.9\% confidence that the average of the model is greater than the average of the delta hedge.  In our view, which of course is biased by our personal risk aversion, this justifies the choice of our agent over the delta hedge.

\subsection{Strengths and limits of the approach}
\label{sec:strengths}
In the previous section, we presented the performance of a TRVO agent where the option to be hedged had the same characteristics as the one on which the agent had been trained.
In this section, we consider the same agents, without any type of \textit{re-training}, and test them on options characterized as follows:
\begin{itemize}
    \item Single in the money (strike 105) option, 60 days maturity.
    \item Single out of the money (strike 95) option, 60 days maturity.
    \item Single at the money option, 60 days maturity, tested on scenarios where the realized volatility is 30\% (to be compared with the training set scenarios having volatility 20\%).
    \item Portfolio of options with different moneyness (strikes from 90 to 110), 60 days maturity.
\end{itemize}
The results of the tests are summarized in Table \ref{tabel}, by using as performance indicators (as in Figure~\ref{fig:frontier}) the difference between the agent and delta hedge p\&l ($\Delta$p\&l) and the difference between the agent and delta hedge p\&l volatility ($\Delta\sigma$).
An optimum (in bold) is identified by looking at the risk aversion-value such that the p\&l gain obtained by varying it, is equal to the volatility increase (i.e., at the optimum the frontier has a tangent line with unitary steepness).
\begin{table}
{\small{
\begin{center}
\setlength{\tabcolsep}{2.5pt}
\begin{tabular}{
c|
cc|
cc|
cc|
cc|
cc|
}
&\multicolumn{2}{|c|}{STD option}&\multicolumn{2}{|c|}{ITM option}&\multicolumn{2}{|c|}{OTM option}&\multicolumn{2}{|c|}{High vol}&\multicolumn{2}{|c|}{Portfolio}\\
[1pt]\hline&&&&& &&&&&  \\[-8pt]
$\lambda$ & $\Delta$p\&l & $\Delta\sigma$ & $\Delta$p\&l & $\Delta\sigma$ & $\Delta$p\&l & $\Delta\sigma$ &$\Delta$p\&l & $\Delta\sigma$ &$\Delta$p\&l & $\Delta\sigma$  \\
0.5 & 0.2 & 0.17 & 0.15 & 0.17 & 0.17 & 0.21 & 0.2 & 0.37 & 0.15 & 0.22 \\
0.75 & 0.19 & 0.12 & 0.14 & 0.13 & 0.15 & 0.15 & 0.19 & 0.23 & 0.14 & 0.14\\
0.82 & 0.18 & 0.1 & 0.13 & 0.1 & 0.15 & 0.12 & 0.18 & 0.2 & 0.13 & 0.12\\
1 & 0.17 & 0.08 & 0.13 & 0.08 & 0.14 & 0.1 & 0.17 & 0.17 & 0.13 & 0.1\\
2 & \textbf{0.14} & \textbf{{0.03}} & \textbf{0.1} & \textbf{0.02} & 0.12 & 0.03 & 0.13 & 0.08 & \textbf{0.1} & \textbf{0.05}\\
3 & 0.11 & 0.01 & 0.08 & 0 & \textbf{0.1} & \textbf{0} & \textbf{0.11} & \textbf{0.04} & 0.08 & 0.03\\
4 & 0.09 & 0 & 0.07 & -0.01 & 0.08 & -0.01 & 0.09 & 0.03 & 0.07 & 0.03\\
5 & 0.08 & -0.01 & 0.06 & -0.02 & 0.07 & -0.01 & 0.07 & 0.02 & 0.06 & 0.02\\
6 & 0.07 & -0.01 & 0.05 & -0.02 & 0.06 & -0.01 & 0.06 & 0.02 & 0.05 & 0.02\\
7 & 0.06 & -0.01 & 0.05 & -0.02 & 0.05 & -0.01 & 0.06 & 0.02 & 0.05 & 0.02\\
8 & 0.06 & -0.01 & 0.04 & -0.02 & 0.05 & -0.01 & 0.05 & 0.01 & 0.04 & 0.02\\
9 & 0.05 & -0.01 & 0.04 & -0.02 & 0.05 & -0.01 & 0.05 & 0.01 & 0.04 & 0.02\\
10 & 0.05 & -0.01 & 0.04 & -0.02 & 0.04 & -0.01 & 0.04 & 0.01 & 0.03 & 0.02
\end{tabular}
\end{center}
}}
\caption{Behavior of agent on a test environment where the option characteristics have been modified.}
\label{tabel}
\end{table}
Table \ref{tabel} proves the robustness of the presented approach: we believe that a single training may be sufficient to properly hedge any portfolio of options for a given maturity.
In fact, considering the state as defined in the accounting formulation (Section \ref{ssec:embedding}), the agent learns a hedging strategy which is independent of the number of options, their strike, and also on the behavior of the market.
Notice also that the $\lambda$ realizing the optimum (with the same considerations as Section \ref{sec:eff_front}) does not change significantly in the table, indicating that a given $\lambda $ consistently realizes a certain risk - p\&l balance, at least for the hedging cost level we adopted.

One could argue that our training/testing is built on a really simple market generation model: GMB with constant volatility; we believe that having a strategy able to deal with a volatility change of around 50\% (from 20\% to 30\%) is already robust and that a real improvement in the strategy would require a massive injection of reality, both on the option and underlying details, and on the market data generation, which could take advantage of AI-based approaches (e.g., the use of a generative adversarial network \cite{NIPS2014_5423}, as in \cite{kondratyev18}, or other approaches such as the Restricted Boltzmann Machine described in \cite{kondratyev19}) in order to be more adherent to real-world data. We leave this extension for future work.

We also tried to understand whether this robustness could be extended to the management of portfolios of options with different maturities. Unfortunately, the first attempts we made seem to indicate that it is extremely difficult to train the agent to learn that, from a certain point on, one or more options expire and so hedging that portion of the portfolio is no longer necessary. In our view, the essential point is that the price/value/delta relationship is badly broken after expiry and even the use of expiry signals to inform the agent about the fact that something changed in the game cannot solve this point completely.
On the other hand, we must observe that for most of the options traded on the market, the available option maturities are not so many, thus it is perfectly viable to split the whole portfolio into maturity sets, each of which is managed by a different instance of the same agent.
\subsection{Managing increasing hedging costs}
\label{sec:experiment_high_cost}
Until now,  we considered hedging costs given by Equation~(\ref{eq:costs}) with $\text{tick size} = 0.05$.
Less liquid or less standard listed contracts may have a significantly  higher tick size (e.g. the Euro Stoxx Banks active future contract, having rescaled tick size $\sim 0.2$), while OTC instruments, such as swaps and CDS, which are perfectly viable option underlyings, have  even higher transaction costs. For this reason we verified what happens with costs with $\text{tick size} = 0.2$.
In such a setting the average cost of the delta hedge is 4-fold (from $\sim 0.286$ to $\sim 1.2$), as well as the cost distribution width. This enhances the advantage of a parsimonious agent: it is possible to draw an efficient frontier in the p\&l/reward volatility space similar to Figure~\ref{fig:frontier0}, just with an increase in the y-axis from 0.2 to 0.8.
On the other hand, a greater width for the cost distribution means a greater p\&l volatility induced on the delta hedge {\it by the hedging costs}, which become the predominant source of volatility.
This effect on p\&l volatility is such that an agent, when reducing hedging costs, may be able to reduce the p\&l volatility {\it as well}.

In this sense, in the presence of higher hedging costs, a winning strategy seems to be {\it decreasing} the risk aversion\footnote{Which does not necessarily imply decreasing the risk aversion parameter $\lambda$, which is not dimensionless.}. In fact, as also mentioned in \cite{cao19}, an optimal strategy in case of very high costs may require no hedging at all. We also tested the (extreme) case of $\text{tick size} = 0.5$, where the described behavior is enhanced even further, essentially recovering the results of \cite{cao19}. As mentioned, the agent is able to outperform delta hedging both in  p\&l and in p\&l volatility, we stress that even in this extreme case the relative outperformance depends on the risk aversion parameter, which in \cite{cao19} was chosen as a fixed parameter ($\lambda \sim 1.5$ using our language).

It would be interesting to understand the precise interplay between risk aversion and hedging costs and volatility and hedging costs in defining an optimal strategy. We leave this for future work, where we believe this kind of issue could be tackled by means of an approach based on Multi Objective Reinforcement Learning~\cite{roijers2013survey}.

\section{Related works}
\label{sec:related}
The issue of automatic hedging has been analyzed by various authors.
Among the most recent approaches we mention \cite{kolm2019dynamic,buehler2019deep,halperin2017qlbs,halperin2019qlbs,cao19}.
These papers can be subdivided into two categories, one addresses the problem from a practitioner's perspective and is focused on the details of the hedging strategies chosen by the agent; the other builds on the formal mathematical structure of option pricing and uses machine learning techniques to overcome the problems posed by realistic features such as transaction costs.
The distinction is faint as a hedging strategy implies a price, and vice-versa.

 The first category includes \cite{kolm2019dynamic,cao19} and is also appropriate for this paper.
 The most comparable to this work regarding the financial environment is~\cite{kolm2019dynamic}, in particular with the use of the accounting formulation as defined in Section~\ref{ssec:embedding}. The main difference consists in the use of a value function RL algorithm. In particular~\cite{kolm2019dynamic} uses a one-step SARSA update which is inspired by Equation~(\ref{eq:bellman}). Using the notation from Section~\ref{sec:RL}, it is defined as: 
\begin{equation*}
    Q(s_t,a_t) \rightarrow \mathcal{R}(s_t,a_t) + \gamma Q(s_{t+1},a_{t+1}),
\end{equation*}
 It is difficult to compare experimental results, as the implementation details are generic. The main comments are that, while in this paper we consider continuous actions, with SARSA or other value function methods it is necessary to descretize the actions and then fit a regressor such as a Neural Network or Random Forest in order to approximate each point. Furthermore, the success of these methods is very sensible to the hyperparameters considered in the regressors. On top of those, there are also the parameters of the SARSA algorithm such as the learning rate $\alpha$, the discount rate $\gamma$, the batch size, and the number of batches. The implementation requires memory enhancements as for each batch it is necessary to store the Q-value of the entire batch for each possible action. This means that multiplying the size of the state, such as the maturity of the option from 10 days (as is in \cite{kolm2019dynamic}) to 60 as in our paper, while keeping the batch size constant in terms of option scenarios, means increasing the data-set size by 6 and thus significantly increasing memory usage.
 Risk aversion is given by only changing the reward to $\mathcal{R}_t = \rho_t-\lambda \rho_t^2$ as mentioned in section~\ref{ssec:riskaverse}.
 In both \cite{kolm2019dynamic, cao19}, only a single value of risk aversion is tested, thus not giving  a frontier but simply showing that hedging is possible.
 
 \cite{cao19} also considers an environment very close to ours, but with a transaction costs size which is $\sim$ 20 times more than what we considered in Section~\ref{sec:experiment_std_cost}. As mentioned in Section~\ref{sec:experiment_high_cost}, extreme hedging costs make the agent very efficient in optimizing both the p\&l and p\&l volatility, but such a feature is not generic and is lost at lower costs.
 Regarding the RL algorithm, they use value-function methods and in particular risk-averse deep Q-learning. It is an advanced approach taken from the risk-averse reinforcement learning literature \cite{tamar2016learning}. They consider two Q functions, one for the first moment and another for the second moment, thus estimating the variance in a more accurate way than in \cite{kolm2019dynamic}, where the second moment is approximated to zero. 
 The paper then focuses on the agent's efficiency as a function of the rebalancing frequency. 

The second category includes \cite{halperin2017qlbs,halperin2019qlbs,buehler2019deep}.
In \cite{halperin2017qlbs,halperin2019qlbs} the problem of option pricing in discrete time has been addressed from a machine learning perspective, neglecting hedging costs, which have been considered in \cite{buehler2019deep}.
In the latter, the option pricing problem is undertaken by considering a class of convex risk measures and embedding them in a deep neural network environment. In the paper, the dependence of the option price and hedge on the risk aversion parameter is deeply studied in the absence of transaction costs. Then a study of the option price dependence on transaction costs is discussed and the functional dependence of the price on the cost parameter is reconstructed. 

\section{Conclusions and Outlook}
\label{sec:conclusions}
In this paper we have shown that it is possible, through the use of risk-averse reinforcement learning, to hedge options ``more optimally'' than what is achieved by delta hedge, which is based on unrealistic assumptions such as continuous rebalancing frequency and no transaction costs.
We used the state-of-the-art algorithm TRVO, which is perfect for this environment as it has a tunable risk-aversion parameter and great learning capability.

In an extensive experimental section, we showed that it is possible to balance risk and return a priori by deciding the agent's level of risk aversion, and how the policies learned are robust as the agents can efficiently hedge options with different characteristics or markets which behave differently than those used in training.

These results have been achieved by considering hedging costs aligned with those that can occur on the market for very liquid assets. In this paper we also considered higher costs, obtaining extremely promising preliminary results. We leave for future work the extension to the management of realistic instruments (such as options on credit indices) where such costs are effectively realized.

Future work could also be to study the relative interplay between rebalancing frequency, risk aversion, and hedging costs. Furthermore, it would be interesting to understand how an approach based on Multi Objective Reinforcement Learning could generate a frontier with a single training.

\bibliographystyle{ACM-Reference-Format}
\bibliography{references}
\end{document}